\documentclass[1p,times,twocolumn]{elsarticle}

 \usepackage{graphics}
 \usepackage{graphicx}
 \usepackage{epsfig}

 \usepackage{color,soul}
\usepackage{amssymb}
\usepackage{textcomp}
\usepackage{gensymb}
\usepackage{booktabs}

\usepackage{enumitem}
\usepackage[version=3]{mhchem} 
\usepackage{esvect} 

 \biboptions{sort&compress}

\journal{Journal of Magnetism and Magnetic Materials}

\begin{document}

\begin{frontmatter}



\title{Magnetic and mechanical effects of Mn substitutions in \ce{AlFe2B2}}


\author[CHEM]{Johan Cedervall\corref{cor1}}\ead{johan.cedervall@kemi.uu.se}
\author[SSP,CH]{Mikael S. Andersson}
\author[PHYS]{Diana Iu\c{s}an}
\author[PHYS]{Erna K. Delczeg-Czirjak}
\author[CHEM]{Ulf Jansson}
\author[SSP]{Per Nordblad}
\author[CHEM]{Martin Sahlberg}

\address[CHEM]{Department of Chemistry - \AA{}ngstr\"{o}m Laboratory, Uppsala University, Box 538, 751 21 Uppsala, Sweden.}
\address[SSP]{Department of Engineering Sciences, Uppsala University, Box 534, 751 21 Uppsala, Sweden.}
\address[CH]{Department of Chemistry and Chemical Engineering, Chalmers University of Technology, G\"{o}teborg 412 96, Sweden}
\address[PHYS]{Division of Materials Theory, Department of Physics and Astronomy, Uppsala University, Box 516, 751 20 Uppsala, Sweden.}

\cortext[cor1]{Corresponding author}

\begin{abstract}
The mechanical and magnetic properties of the newly discovered MAB-phase class of materials based upon \ce{AlFe2B2} were investigated. The samples were synthesised from stoichiometric amounts of all constituent elements. X-ray diffraction shows that the main phase is orthorhombic with an elongated $b$-axis, similar to \ce{AlFe2B2}. The low hardness and visual inspection of the samples after deformation indicate that these compounds are deformed via a delamination process. When substituting iron in \ce{AlFe2B2} with manganese, the magnetism in the system goes from being ferro- to antiferromagnetic via a disordered ferrimagnetic phase exhibited by \ce{AlFeMnB2}. Density functional theory calculations indicate a weakening of the magnetic interactions among the transitions metal ions as iron is substituted by manganese in \ce{AlFe2B2}. The Mn-Mn exchange interactions in \ce{AlMn2B2} are found to be very small.
\end{abstract}


\end{frontmatter}


\section{Introduction}
The discovery of the giant magnetocaloric effect (GMCE) in \ce{Gd5(Si2Ge2)} in 1997~\cite{Pecharsky1997} started off an intense search for materials to be used for magnetic refrigeration. A refrigerator based on the GMCE has the theoretical possibility to become 20-30\% more effective than a conventional vapour compression refrigerator~\cite{GschneidnerJr2008}. This could, therefore, contribute to a more sustainable future, especially since predictions show that there is going to be an increased demand for refrigeration in the future~\cite{Gauss2016}. For materials showing GMCE, the most sustainable ones would be those which contain cheap and abundant elements; this excludes usage of the rare earth elements~\cite{Coey2012}. Among the proposed candidates, some of the most researched material are \ce{(Fe{,}Mn)2(P{,}Si)}, based on the \ce{Fe2P} structure~\cite{Hoglin2015, Miao2014}, \ce{La(Fe{,}Si)13}, and Heusler-based alloys~\cite{Bruck2005}. All these have shown Curie transition temperatures (T$\rm _C$) that may be tuned to around room temperature, as well as high magnetic entropy changes ($\Delta \rm S_{mag}$).
	
Another recently studied material system for magnetic cooling is \ce{Al\it{M}2B2} ($M$~=~Fe,~Mn,~Cr), based on \ce{AlFe2B2}~\cite{Jeitschko1969, ElMassalami2011, Tan2013, Chai2015, Lewis2015, Cedervall2016a, Cedervall2016b, Ali2017}. The structure of \ce{AlFe2B2} is orthorhombic (space group $Cmmm$) with \ce{(Fe2B2)}-slabs in between layers of aluminium stacked along the $b$-axis~\cite{Jeitschko1969}. The unit cell contains two formula units with the aluminium atoms occupying the $2a$ position and the iron and boron atoms occupying the $4i$ and $4j$ positions, respectively. The unit cell parameters are 2.9233(10), 11.0337(14), and 2.8703(3)~\AA{} for $a$, $b$, and $c$, respectively. Previous studies have shown that the compound is ferromagnetic with an ordering temperature (T$\rm _C$) around 300~K and magnetic moments close to 1~$\mu _B$/Fe-atom~\cite{ElMassalami2011, Tan2013, Chai2015}. A recent neutron diffraction study confirmed that the compound is ferromagnetic with the magnetic moments aligned along the crystallographic $a$-axis~\cite{Cedervall2016a}. The same study also revealed that the magnetic transition is of second order, which makes it less suitable for application in magnetic refrigeration. The second order magnetic phase transition has also been confirmed by other studies~\cite{Ali2017}. A study performed on elongated crystals confirmed that the easy axis is along the $a$-axis~\cite{Barua2018}. Additionally, the presence of magnetocrystalline anisotropy leads to a substantial rotational entropy change, $\Delta \rm S_{rot}$, in \ce{AlFe2B2}. A theoretical study revealed a calculated relative cooling power (RCP) of 480~J/kg, which exceeds the reported experimental values and increases the potential use of the \ce{Al\it{M}2B2} compounds in applications~\cite{Boukili2018}. Electronic structure calculations for \ce{Al\it{M}2B2} with a mixed metal occupation suggest an antiferromagnetic structure for \ce{AlMn2B2} and that the characteristics on the magnetic ordering change when when going from \ce{AlFe2B2} to \ce{AlMn2B2}~\cite{Ke2017}. Performing substitutions on the Al-site in \ce{AlFe2B2} with Ga and Ge has also shown a tunability of the magnetocaloric response~\cite{Barua2019}.
	
The crystal structures of \ce{AlMn2B2} and \ce{AlCr2B2} have been known for several decades and are isostructural with \ce{AlFe2B2}~\cite{Becher1966, Chaban1973}. Partial substitution on the transition metal site in \ce{AlFe2B2} can tune the magnetic properties, where 60\% manganese substitution of iron is reported to lower the T$\rm _C$ down to 43~K, while pure \ce{AlMn2B2} is suggested to be antiferromagnetic~\cite{Chai2015}. This was confirmed using neutron diffraction where a canted antiferromagnetic ordering (below 390~K) was found with the magnetic moments oriented either between the $a$ or $c$-axes with a doubling of the crystallographic unit cell in the $c$-direction~\cite{Potashnikov2019}. By substituting iron on the transition metal site in \ce{AlFe2B2} with cobalt, the T$\rm _C$ decreases linearly with increasing cobalt concentration, down to 205~K for 30\% substitution~\cite{Hirt2016}. $\Delta \rm S_{mag}$ did not, however, change significantly with the substitution.
	
It was also suggested that the \ce{Al\it{M}2B2}-compounds belong to a new class of materials called MAB-phases~\cite{Ade2015}. The MAB-phases are similar to MAX-phases where transition metal-carbides or nitrides are stacked between aluminium layers. The \ce{Al\it{M}2B2} phase can delaminate due to weak chemical bonds between the (\ce{\it{M}2B2})-slabs and Al. However, in contrast to the MAX-phases, no kink bands have yet been observed fo \ce{Al\it{M}2B2}~\cite{Kadas2017}. This is also shown in the hardness values which were found to be 10.4(3), 7.3(3), and 9.5(3)~GPa for \ce{AlCr2B2}, \ce{AlMn2B2}, and \ce{AlFe2B2} respectively, which is lower than usual for metal borides (typically around 20-30~GPa~\cite{Holleck1986}). For \ce{AlFe2B2}, other mechanical properties, such as compressive strength and fracture toughness, were reported to be 2.1(2)~GPa and 5.4(2)~MPa$\cdot$m$^{1/2}$, respectively~\cite{Li2017}.
	
The present study reports on the synthesis of stoichiometric aluminium metal borides with a mixed occupancy on the metal site. It includes the magnetic and mechanical properties of a mix of iron and manganese in this system and focuses on the changes that occur from the substitutions. We have investigated this via X-ray diffraction, magnetic measurements, electronic structure calculations, hardness testing, as well as studies of the delamination process.
	
\section{Experimental and computational details}
Samples of the compositions \ce{AlFe2B2}, \ce{AlFeMnB2}, and \ce{AlMn2B2} were synthesised with stoichiometric amounts of pure elements. The elements aluminium (Gr\"anges SM, purity 99.999\%), iron (Leico Industries, purity 99.99+\%. Surface oxides were reduced in \ce{H2}-gas.), manganese (Institute of Physics, Polish Academy of Sciences, purity 99.999\%), and boron (Wacher-Chemie, purity 99.995\%) were placed in an arc furnace and were melted five times with flipping of the pieces between each melting to ensure maximum homogeneity. All samples were crushed, and the obtained pieces were placed in evacuated silica ampoules and annealed for 14~days at 1173~K and subsequently quenched in cold water.
	
The crystalline phase content of all samples was analysed with X-ray powder diffraction (XRD) which was performed with a Bruker D8 diffractometer equipped with a Lynx-eye position sensitive detector (PSD, 4\textdegree\ opening) using CuK$\alpha_1$ radiation ($\lambda$~=~1.540598~\AA ), in a 2$\theta$ range of 20-90\textdegree\ at room temperature. The obtained XRD patterns were also used for analysis of the crystalline structures utilising the Rietveld method~\cite{Rietveld1969}, which was performed with the FullProf software~\cite{Rodriguez-Carvajal1993}. The unit cell parameters were determined with refinements performed with the software UnitCell~\cite{Holland1997}. 

Thermal analyses were performed on powdered samples using a Netzsch STA 409 PC instrument at a heating rate of 10~K/min under Ar flow up to 1570~K. In order to assess the mechanical properties, the annealed samples were placed in bakelite and polished to a roughness of \textless 1~$\mu$m. The hardness was then evaluated via Vickers micro-hardness measurements using a Matsuzawa MTX50 with a load of 200~g dwelling for 15~s. The measurements were done 10~times in the same region on each sample, and the mean value for the hardness is reported here. Polished samples were split into separate pieces by the application of a large force to the surface using a sharp diamond tip. The delamination at the cleavage surface of the samples were studied with a FEG-Zeiss Merlin scanning electron microscopy (SEM) equipped with an AZtec energy dispersive X-ray spectrometer (EDS) and Electron Backscatter Diffraction (EBSD) detector. 
	
Magnetisation measurements as a function of temperature and as a function of an applied magnetic field were performed using an MPMS SQUID magnetometer from Quantum Design. The measurements for the magnetisation as a function of temperature were performed using magnetic fields $\mu_0H$~=~1~T (10~000~G) and $\mu_0H$~=~5~mT (50~G) over the temperature range 10-400~K. The measurements for the magnetisation as a function of the applied magnetic field were performed by sweeping the field between $\mu_0H$~=~$\pm$5~T ($\pm$50 000~G).

The experimental measurements have been complemented by electronic structure calculations within density functional theory (DFT). For this purpose, the full-potential linear muffin-tin orbital (FP-LMTO) method as implemented in the RSPt code have been employed~\cite{Wills:2010ej}. The local density approximation for the exchange-correlation functional was chosen. The maximum value of the angular momentum $(l)$ for the expansion of the potential and the electron density within the muffin-tin spheres was $l_{max} = 12$. Three kinetic energies were used for the basis in the interstitial region: -0.1, -2.3, and -0.6~Ry. The Brillouin zone integration was performed on a $32\times32\times32$~{\bf k}-points mesh.

Using this setup the Heisenberg exchange parameters among the magnetic atoms via the Liechtenstein-Katsnelson-Antropov-Gubanov (LKAG) formalism were calculated~\cite{Liechtenstein:1984fj, Liechtenstein:1987br}. The magnetic Hamiltonian is
\begin{equation}
	H = \sum_{i \ne j} J_{ij}~\vec{e_i} \cdot \vec{e_j},
	\label{eq:spinH}
\end{equation}
where $\vec{e_i}$ denotes the unit vector along the direction of the magnetic moment at site $i$. For the particular implementation of the LKAG formalism into the RSPt code, the reader is refered to Ref.~\cite{Kvashnin:2015fx}. The chemical disorder at the transition metal sites was treated within the virtual crystal approximation (VCA). 

\section{Results and discussion}
\begin{figure}[t]
\centering
\includegraphics[width = 0.49\textwidth]{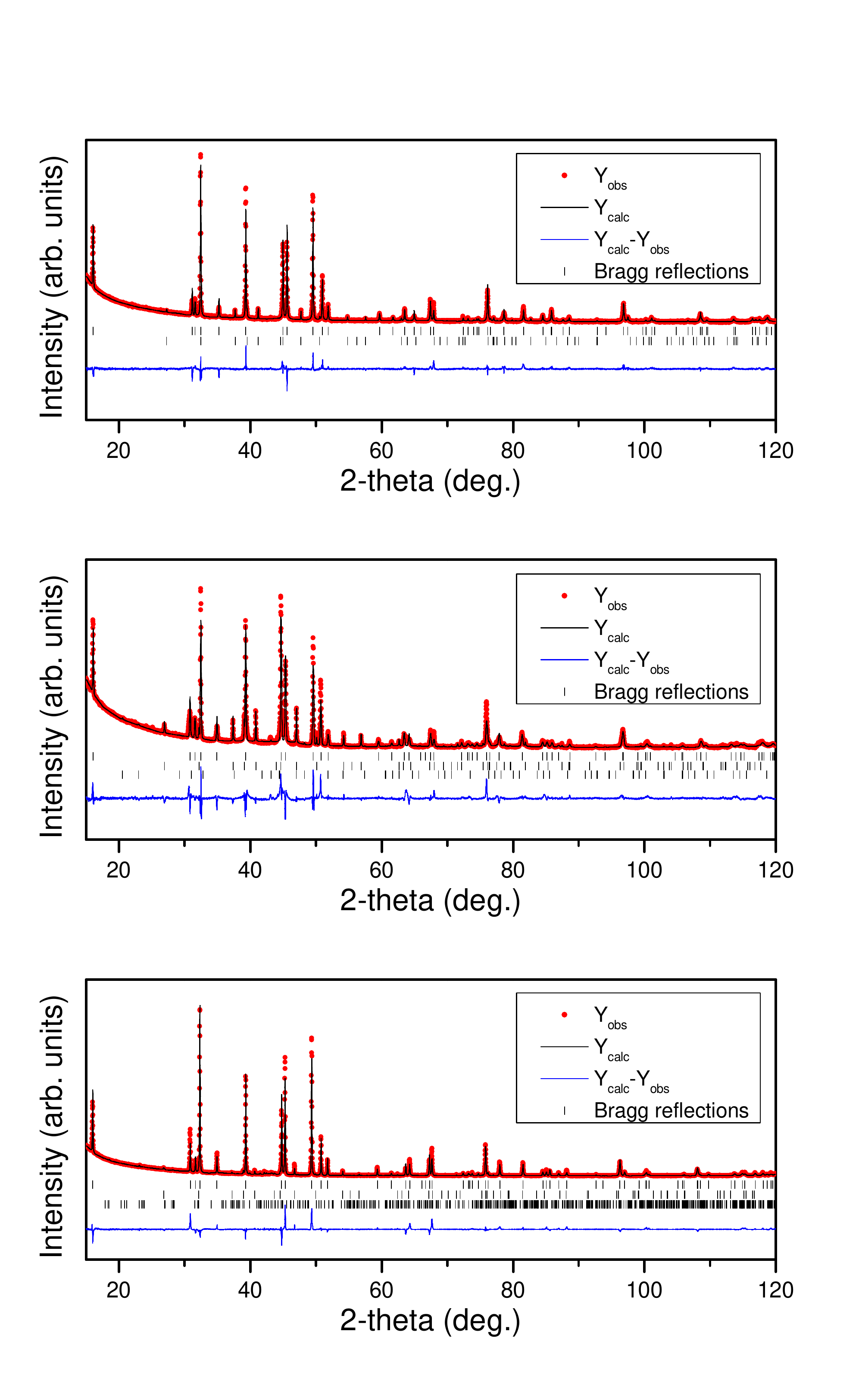}
\caption{Refined X-ray diffraction patterns for \ce{AlFe2B2} (top), \ce{AlFeMnB2} (middle), and \ce{AlMn2B2} (bottom). $\lambda$~=~1.540598~\AA. The phases present in each sample (shown as tick marks in the figure) are, from top to bottom, \ce{Al\it{M}2B2} and \ce{FeB} for \ce{AlFe2B2}, \ce{Al\it{M}2B2}, \ce{MnB} and \ce{Fe3B} for \ce{AlFeMnB2} and \ce{Al\it{M}2B2}, \ce{MnB} and \ce{Mn4Al11} for \ce{AlMn2B2}.}
\label{AlM2B2-XRD}
\end{figure}

The refined XRD patterns for the \ce{Al\it{M}2B2}-samples are shown in Figure~\ref{AlM2B2-XRD} and indicate that the main phase for each of the tree compounds crystallises in the layered orthorhombic structure with the space group $Cmmm$. The synthesis procedure with stoichiometric amounts of all elements results in multiphase samples. The secondary phases in the samples are, for \ce{AlFe2B2}, \ce{FeB} (\textless 2\%), for \ce{AlFeMnB2}, \ce{MnB} (\textless 6\%) and \ce{Fe3B} (\textless 1\%), and for \ce{AlMn2B2}, \ce{MnB} (\textless 1\%) and \ce{Mn4Al11} (\textless 1\%). Table~\ref{Tab:Unitcell} summarises the refined unit cell parameters, which are in agreement to the values previously reported for \ce{AlMn2B2}~\cite{Becher1966} and \ce{AlFe2B2}~\cite{Jeitschko1969}. \ce{AlFeMnB2} has unit cell parameters similar to the other compounds, which is reasonable given the similar atomic radius for iron and manganese. There is a small difference between the unit cell parameters observed here (and previously by us~\cite{Cedervall2016a, Kadas2017}) and the parameters by Chai $\textit{et al.}$~\cite{Chai2015}, which could be an indication of a possible homogeneity range of the \ce{Al\it{M}2B2} phase. We have, however, prepared samples both in stoichiometric ratios between the elements (here) as well as with excess Al (\cite{Cedervall2016a, Kadas2017}) and these values are in close agreement with each other, suggesting that this is not the case. Using the above described synthesis route attempts of  chromium substitution  were made. However, the desired phase only occurred as a minority phase, implying that an alternative synthesis route has to be used for successfully substituting iron for chromium.

\begin{table*}[b]
\small
\caption{Refined unit cell parameters for \ce{Al\it{M}2B2}. Standard deviations are given in the parenthesis.}
\begin{tabular*}{\textwidth}{@{\extracolsep{\fill}}lllll}
\toprule
\ce{Al\it{M}2B2}	& a (\AA) 		& b (\AA)		& c (\AA)		& V (\AA$^3$)		\\
\midrule 
\ce{AlFe2B2} 	&  2.9263(3)	& 11.0295(9)	& 2.8666(3)	&	92.52(2)		\\
\ce{AlFeMnB2} 	&  2.9206(2)	& 11.0673(7)	& 2.8957(3)	&	93.60(1)		\\
\ce{AlMn2B2}	&  2.9300(6)	& 11.0186(12)	& 2.8975(8)	&	93.54(3)		\\
\bottomrule
\label{Tab:Unitcell}
\end{tabular*}
\end{table*}
	
\begin{figure}[ht]
\centering
\includegraphics[width = 0.49\textwidth]{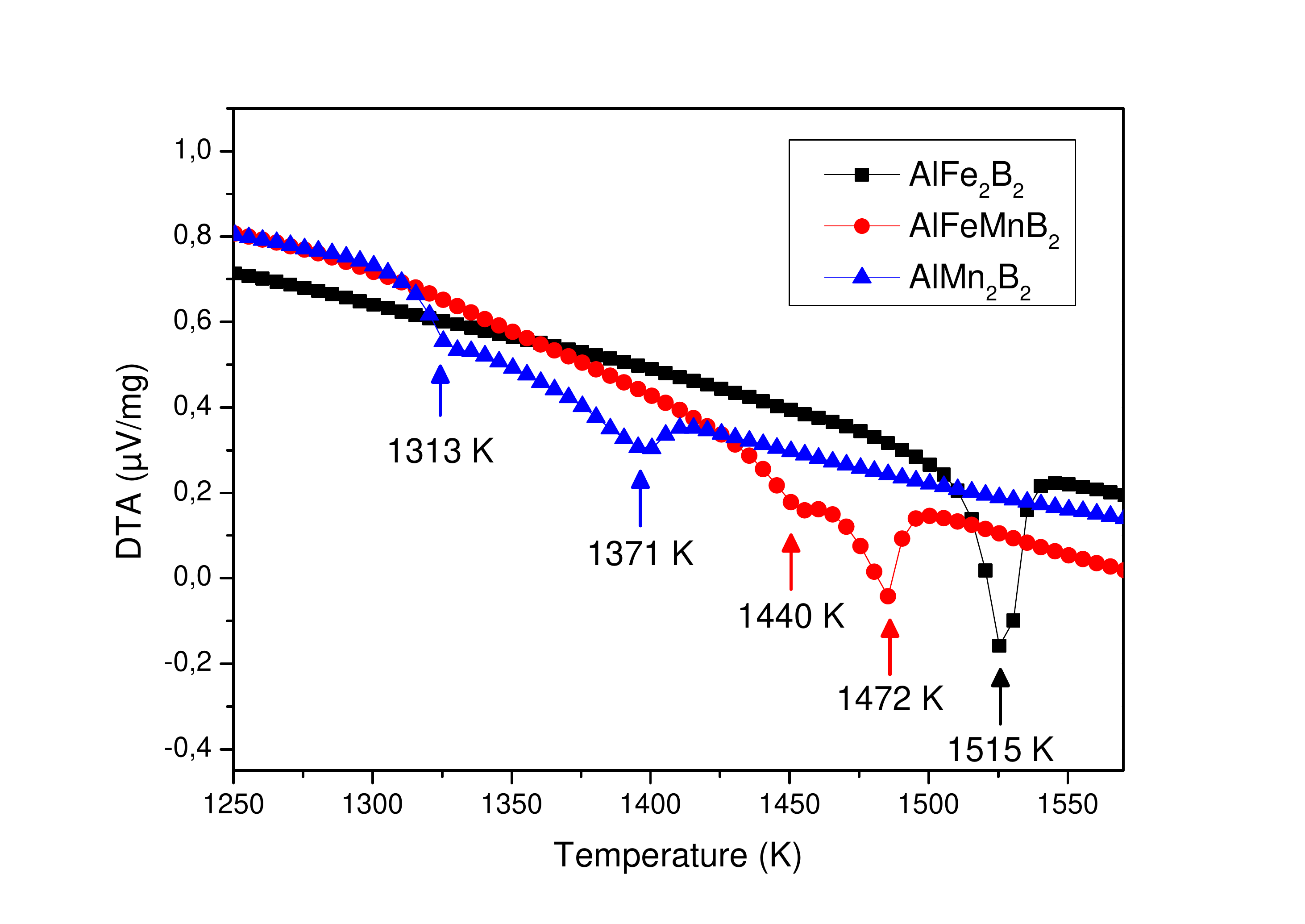}
\caption{DTA curves for \ce{AlFe2B2}, \ce{AlFeMnB2} and \ce{AlMn2B2}. Onset temperatures of the endothermic changes are indicated.}
\label{AlM2B2-DTA}
\end{figure}

The differential thermal analyses (DTA) performed on \ce{AlFe2B2}, \ce{AlFeMnB2}, and \ce{AlMn2B2}, presented in Figure~\ref{AlM2B2-DTA}, show the melting points of all compounds. It is clear that \ce{AlFe2B2} melts at the highest temperature (1515~K) and that the melting point is lowered with increasing manganese substitution, 1472 and 1371~K for \ce{AlFeMnB2} and \ce{AlMn2B2}, respectively. For both \ce{AlMn2B2} and \ce{AlFeMnB2}, two peaks appear upon melting, which could potentially come from secondary phases. Nevertheless, it is more likely that this is due to incongruent melting, which can also be presumed from the difficulties in making samples when transition metals other than iron are present in the synthesis.
	
\begin{figure*}[b]
\centering
\includegraphics[width = 0.99\textwidth]{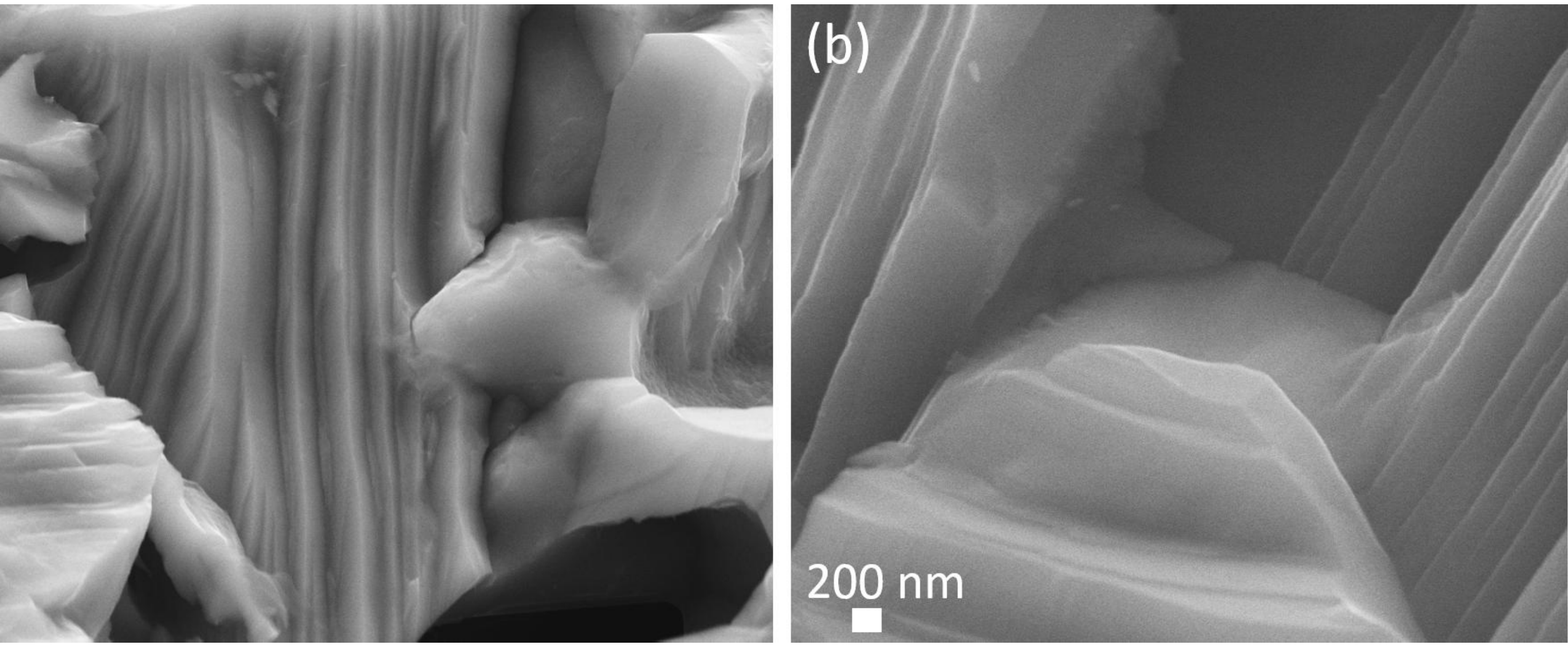}
\caption {SEM micrographs with delamination from indents in different directions for \ce{AlFeMnB2}.}
\label{AlM2B2-SEM}
\end{figure*}

The measured hardness values are 9.3(3), 10.6(2), and 7.2(3)~GPa for \ce{AlFe2B2}, \ce{AlFeMnB2}, and \ce{AlMn2B2}, respectively. This is consistent with the trend for the calculated ratio of bulk moduli over shear moduli~\cite{Kadas2017}. Whether the differences in hardness are an effect of the substitutions or coming from precipitations of secondary phases is not easily judged. Nevertheless, the values are significantly lower than normally found for metal borides~\cite{Holleck1986}, which can be attributed to the nanolaminated structure of these phases with weak bonds between the (\ce{\it{M}2B2}) and Al layers~\cite{Ade2015, Kadas2017}. In figure~\ref{AlM2B2-SEM} micrographs for the \ce{AlFeMnB2} sample is shown, where indents with a diamond tip were made. The indents show similar behaviour as previously reported and it was suggested to come from delamination between the layers in the structure. The same reasoning can therefore be applied here to explain the low hardness of these materials. EDS measurements on the \ce{AlFeMnB2}-sample shows that iron and manganese are almost equally distributed. The small mismatch between the manganese and iron contents in the samples might come from precipitates of secondary phases. The mismatch also indicates that grains with different Fe/Mn-contents exist, with slightly different unit cell parameters. This can also be observed given the small misfit in the refined XRD pattern in Figure~\ref{AlM2B2-XRD}. That the compound is not congruently melting can also be seen in the SEM micrographs, and is also suggested from the observed secondary phases. However, the almost equal distribution of the elements further indicates that the substitution is possible.

\begin{figure*}[h]
\centering
\includegraphics[width = 0.99\textwidth]{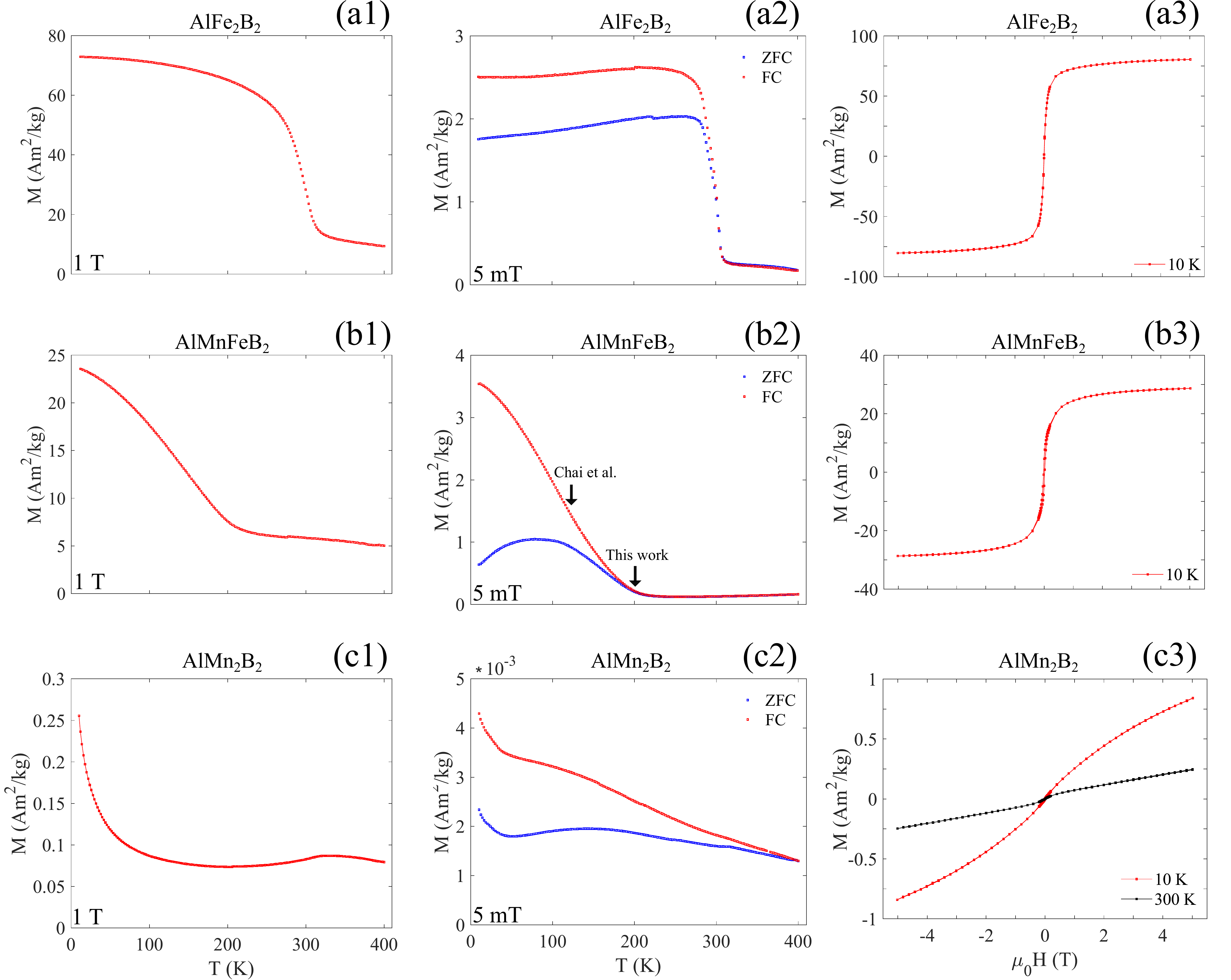}
\caption{Magnetisation as a function of temperature and applied magnetic field. In column \textbf{(x1)}, x = a, b or c, the magnetisation vs. temperature under an applied magnetic field of $\mu_0H$~=~1~T is presented for the three samples. In column \textbf{(x2)} the low field ($\mu_0H$~=~5~mT) magnetisation vs. temperature is shown. In column \textbf{(x3)} the magnetisation vs. an applied magnetic field at 10~K. \textbf{(c3)} also includes a measurement at 300~K. The arrows in (b2) indicate T$_{\rm C}$ found in this work as well as T$_{\rm C}$ found by Chai $\textit{et al.}$ \cite{Chai2015}.}
\label{AlM2B2-Magnetic}
\end{figure*}

The magnetometry measurements as a function of temperature and as a function of applied magnetic field for \ce{AlFe2B2}, \ce{AlFeMnB2}, and \ce{AlMn2B2} are presented in Figure~\ref{AlM2B2-Magnetic}. The results indicate that \ce{AlFe2B2} is a ferromagnet with a T$_{\rm C}$ of about 300~K. The \ce{FeB} impurity phase contributes to the measured magnetisation above T$_{\rm C}$ of \ce{AlFe2B2}. The measured saturation magnetisation at 10~K agrees within error bars with previous findings~\cite{Cedervall2016a, Chai2015, Lewis2015}. Substituting 50\% of the iron with manganese introduces competing ferro- and antiferromagnetic interaction and leads to a decrease in the saturation magnetisation as well as a decrease in T$_{\rm C}$. The decrease in the saturation magnetisation and the temperature dependence of the low field magnetisation near T$_{\rm C}$ at about 200~K for \ce{AlFeMnB2} imply that this system is a frustrated ferrimagnet. For \ce{AlFeMnB2}, the observed magnetisation vs. field and temperature are essentially similar to those reported previously~\cite{Chai2015}. However, the Curie temperature presented here is derived differently compared to previously reported values~\cite{Chai2015}, see Figure~\ref{AlM2B2-Magnetic} (b2). Here T$_{\rm C}$ was defined as the temperature for which the magnetisation starts to increase rapidly. The broad maximum slightly above 300~K in the magnetisation as a function of temperature plot for \ce{AlMn2B2}, as well as a very low magnetisation, indicate that \ce{AlMn2B2} is an antiferromagnet with a T$_{\rm N}$ of about 300~K. This result differs from that obtained by Chai $\textit{et al.}$ who finds a possible antiferromagnetic transition at $\sim$50~K. However, the data presented by Chai $\textit{et al.}$ is limited to a maximum temperature of 300~K, and it is therefore not possible to see a transition occurring above 300~K~\cite{Chai2015}. The density of states calculated for \ce{AlMn2B2} by Chai $\textit{et al.}$ suggests a non-ferromagnetic ground state due to the location of the Fermi level at the lower edge of the antibonding region~\cite{Chai2015}. The results presented here are, however, not contradictory to that, since an antiferromagnetic configuration is, indeed, non-ferromagnetic. From the low field magnetisation as a function of temperature measurements it is clear that the sample contains a magnetic impurity phase with a transition temperature larger than 400~K. This impurity phase can also be seen in the magnetisation as a function of applied magnetic field curves in Figure~\ref{AlM2B2-Magnetic} (c3). From the XRD results this impurity can be identified as \ce{MnB}, T$_{\rm C}$~$\approx$~570~K.

\begin{table*}[b]
\small
\caption{Calculated magnetic moments of the \ce{Al\it{M}2B2} compounds, total value per formula unit, as well as projected onto the transition metal sites. The induced magnetic moments on the Al and B sites are negligible.}
\begin{tabular*}{\textwidth}{@{\extracolsep{\fill}}lllll}
\toprule
\ce{Al\it{M}2B2}	& $\rm \mu_{tot}/f.u.~(\mu_{\rm B}$)	& $\rm \mu_{Fe}~(\mu_{\rm B}$)	& $\rm \mu_{Mn}~(\mu_{\rm B}$)	& $\rm \mu_{Fe_{0.5}Mn_{0.5}}~(\mu_{\rm B}$)	\\
\midrule 
\ce{AlFe2B2}			& 2.47																& 1.27													& -															& -																						\\
\ce{AlFeMnB2} (ordered)			 & 1.63											& 1.17													& 0.53							 & -																											\\
\ce{AlFeMnB2} (disordered)	 & 1.82											& -															& -									 & 0.97																										\\
\ce{AlMn2B2}								 & 0.75		 									& -															& 0.40													& -																						\\
\bottomrule
\label{table:DFTmoments}
\end{tabular*}
\end{table*}

\begin{figure*}[h]
\centering
\includegraphics[width = 0.32\textwidth]{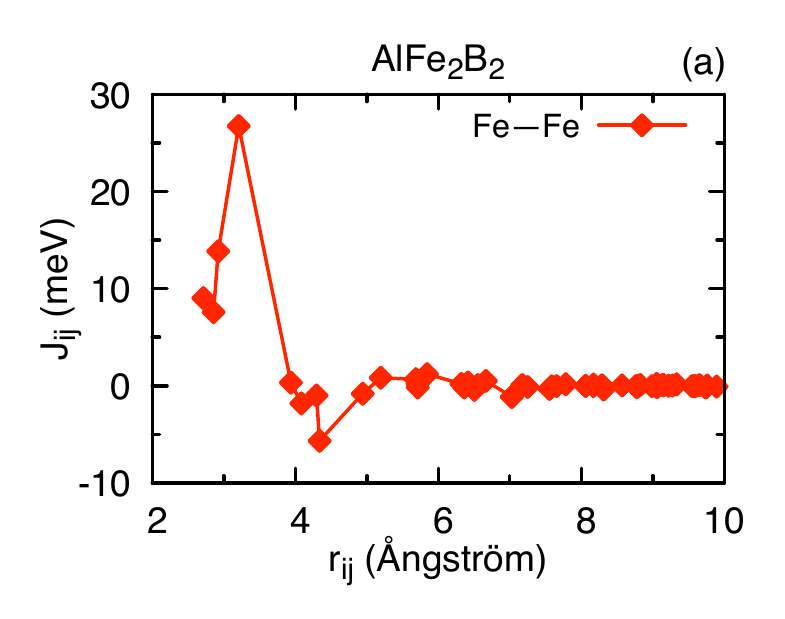}
\includegraphics[width = 0.32\textwidth]{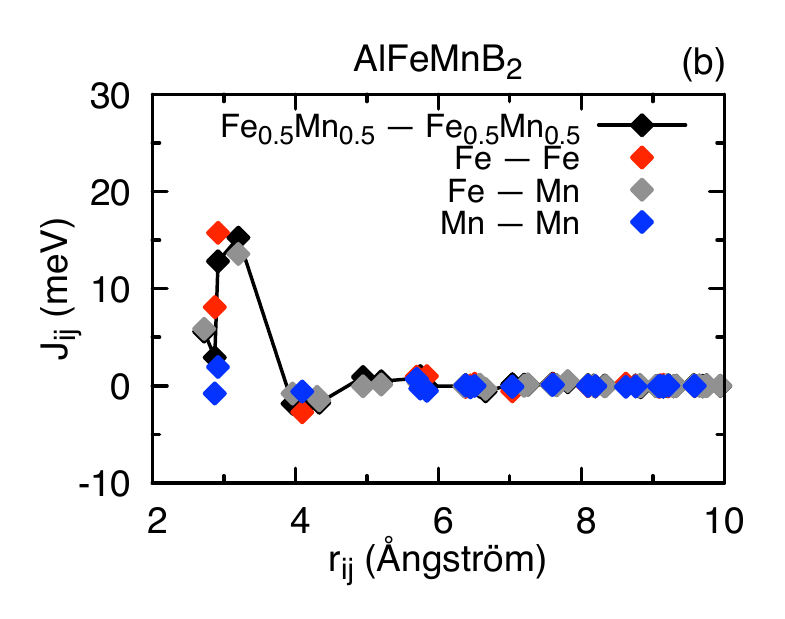}
\includegraphics[width = 0.32\textwidth]{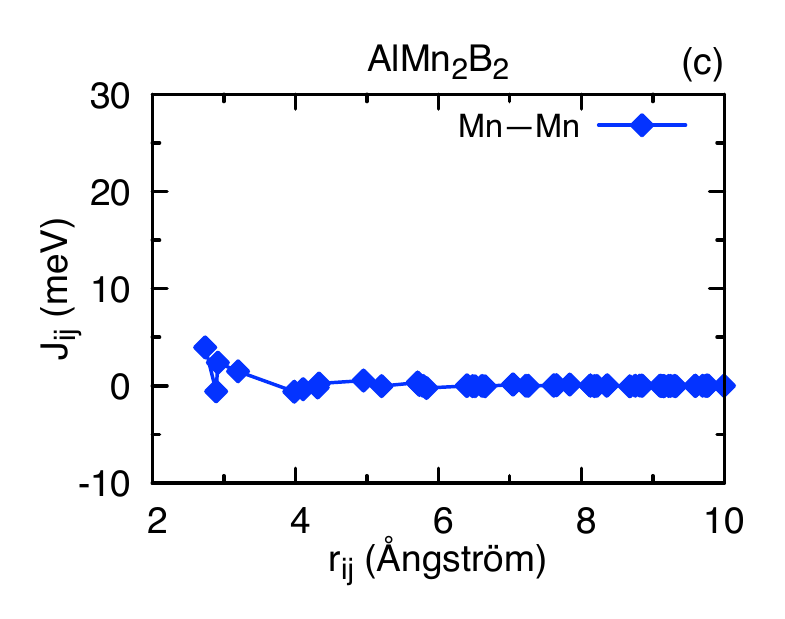}
\caption{Calculated Heisenberg exchange interactions among the transition metal spins at sites $i$ and $j$ as a function of the distance between the spins, $r_{ij}$ in the (a) \ce{AlFe2B2}, (b) \ce{AlFeMnB2}, and (c) \ce{AlMn2B2} compounds. A positive value denotes a ferromagnetic coupling among the spins, while a negative value denotes an antiferromagnetic one. In (b) the calculated exchange interactions in \ce{AlFeMnB2} for both the chemically ordered (colored symbols) and disordered phases (black line and points) are shown.}
\label{figure:AlM2B2-Jijs}
\end{figure*}

The calculated magnetic moments by means of the FP-LMTO method are given in Table~\ref{table:DFTmoments}. For \ce{AlFe2B2}, a total magnetic moment of 2.47 $\mu_{\rm B}{\rm /f.u.}$ was obtained. This value agrees very well with the calculated value of 2.5~$\mu_{\rm B}{\rm /f.u.}$ by Ke $\textit{et al.}$~\cite{Ke2017} and the measured value of 2.3~$\mu_{\rm B}{\rm /f.u.}$, calculated from saturation magnetisation value in Figure~\ref{AlM2B2-Magnetic}~(a3). For \ce{AlMn2B2}, a value of 0.67~$\mu_{\rm B}{\rm /f.u.}$ was obtained in the ferromagnetic configuration, which collapses to zero in the antiferromagnetic one. As for the \ce{AlFeMnB2} compound, both the chemically ordered and disordered configurations were considered. In the chemically ordered phase, one of the transition metal sites is occupied by Fe and the other one by Mn. In the disordered phase, the two transition metal sites have the same average occupation of 50\% Fe and 50\% Mn. The total magnetic moments are similar in the two phases, equal to 1.63 and 1.82~$\mu_{\rm B}{\rm /f.u.}$, respectively. The projected magnetic moments at the Fe- and Mn-sites, 1.17 and 0.53~$\mu_{\rm B}$, respectively, are close to their respective values in \ce{AlFe2B2} and \ce{AlMn2B2}. This may indicate that chemical disorder does not have a critical influence on the magnetism of the \ce{AlFeMnB2} compound.

Figure~\ref{figure:AlM2B2-Jijs} shows the Heisenberg pair-interactions among the magnetic spins, calculated via the Liechtenstein-Katsnelson-Antropov-Gubanov (LKAG) formalism, as defined in Eq.~\ref{eq:spinH}~\cite{Liechtenstein:1984fj, Liechtenstein:1987br}. For \ce{AlFe2B2} ferromagnetic coupling for small Fe-Fe separations and an oscillatory RKKY-like dependence at larger distances were found. The magnetic exchange interactions among the Mn spins in \ce{AlMn2B2} are negligible in size. However, the first nearest neighbour interactions are positive while the next nearest neighbour J$_{ij}$ is negative, which could be indicative of an antiferromagnetic ordering suggested here and in previous experimental work~\cite{Potashnikov2019}. The strength of the exchange interactions in the \ce{AlFeMnB2} compound is approximately half compared to the value in \ce{AlFe2B2}. Their values are rather similar for the chemically ordered and disordered configurations, as observed for the magnetic moments as well. The Mn-Mn interactions in the ordered phase are very small, as it was the case for the \ce{AlMn2B2} compound.

\section{Conclusions}
The synthesis routes presented here give crystalline samples with an orthorhombic phase. However, there exists secondary phases that other synthesis methods for at least \ce{AlFe2B2} have proven to avoid. However, this synthesis route is efficient and closer to what can potentially be achieved with industrial methods. From the mechanical tests presented here, it can be concluded that manganese substitution has a limited effect on the mechanical properties. Magnetometry on the other hand, indicates that going from \ce{AlFe2B2} via \ce{AlFeMnB2} to \ce{AlMn2B2} the system goes from being a ferromagnet to a disordered ferrimagnet and finally to an antiferromagnet, and with a decrease in the saturation magnetisation reflected by this change. 
	
\section*{Acknowledgments}
This work was financed by the Swedish Research Council and the Swedish Energy Agency, which are gratefully acknowledged. The authors also thank Samrand Shafeie for his help with the SEM analysis and Pedro Berastegui for his help with the thermal analysis. D.~I. and E.K.~D.-Cz. would like to acknowledge the Swedish National Infrastructure for Computing (SNIC) for computational resources (project numbers 2017/11-55 and 2018/1-34).




\bibliographystyle{model1a-num-names}
\bibliography{references}







\end{document}